\documentclass[preprint,aps]{revtex4}
\usepackage{graphicx}
                       
\newcommand{\bea}{\begin{eqnarray}}
\newcommand{\eea}{\end{eqnarray}}
\newcommand{\be}{\begin{equation}}
\newcommand{\ee}{\end{equation}}                                             
\newcommand{\Tr}{\mathrm{Tr}}
  
\begin{document}
\title{Meson current in the CFL phase}
\author{A.~Gerhold and T.~Sch\"afer}
\affiliation{Department of Physics\\
North Carolina State University,
Raleigh, NC 27695}

\begin{abstract}
We study the stability of the color-flavor locked (CFL) phase
of dense quark matter with regard to the formation of a non-zero
Goldstone boson current. We show that an instability appears in 
the vicinity of the point $\mu_s=\Delta$ which marks the appearance
of gapless fermion modes in the CFL phase. Here, $\mu_s=m_s^2/(2\mu)$
is the shift in chemical potential due to the strange quark mass
and $\Delta$ is the gap in the chiral limit. We show that in 
the Goldstone boson current phase all components of the magnetic 
screening mass are real. In this work we do not take into account
homogeneous kaon condensation. We study the effects of an instanton
induced interaction of the magnitude required to suppress kaon 
condensation.

\end{abstract}
\maketitle

%%%%%%%%%%%%%%%%%%%%%%%%%%%%%%%%%%%%%%%%%%%%%%%%%%%%%%%%%%%%%%%%%%%%%%%%%
\section{Introduction}
\label{sec_intro}
%%%%%%%%%%%%%%%%%%%%%%%%%%%%%%%%%%%%%%%%%%%%%%%%%%%%%%%%%%%%%%%%%%%%%%%%%

 Exploring the phase diagram of dense baryonic matter is an area of 
significant theoretical and experimental activity. Using methods based 
on weak-coupling QCD it was shown that the groundstate of three flavor 
quark matter at very high baryon density is the color-flavor-locked (CFL) 
phase \cite{Alford:1999mk,Schafer:1999fe,Evans:1999at}. Finding the
second, third, etc.~most dense phase turns out be a much harder 
problem. If the quarks are degenerate and flavor symmetry is exact 
then the symmetries of the CFL state match the symmetries of (hyper) 
nuclear matter. This means that high density quark matter and low
density nuclear matter might be continuously connected, without an 
intervening phase transition \cite{Schafer:1998ef}.

 In the real world the strange quark mass is not equal to the masses
of the up and down quark and flavor symmetry is broken. At high baryon
density the effect of the quark masses is governed by the shift 
$\mu_q = m_q^2/(2\mu)$ of the Fermi energy due to the quark mass. 
This implies that flavor symmetry breaking becomes more important as 
the density is lowered. Clearly, the most important effect is due
to the strange quark mass. There are two scales that are important
in relation to $\mu_s$. The first is the mass of the lightest
strange Goldstone boson, the kaon, and the second is the gap $\Delta$ 
for fermionic excitations. When $\mu_s$ becomes equal to $m_K$
the CFL phase undergoes a transition to a phase with kaon condensation
\cite{Bedaque:2001je}. In the present work we wish to study possible
phase transitions in the vicinity of the point $\mu_s=\Delta$.

 Alford, Kouvaris and Rajagopal observed that gapless fermions appear 
in the spectrum if $\mu_s>\Delta$ and that the CFL state is energetically 
stable compared to other homogeneous phases as long as  $\mu_s<2\Delta$. 
They termed the phase in the regime $\Delta<\mu_s<2\Delta$ the gapless CFL 
(gCFL) phase \cite{Alford:2003fq,Alford:2004hz}. The problem is that gapless
fermion modes in weakly coupled superfluids tend to cause instabilities
in current-current correlation functions \cite{Wu:2003,Huang:2004bg}
and these instabilities have indeed been found in the gCFL phase
\cite{Casalbuoni:2004tb,Alford:2005qw,Fukushima:2005cm}. We have 
suggested that the instability is resolved by the formation of a 
non-zero Goldstone boson current \cite{Kryjevski:2005qq,Schafer:2005ym}.
We will refer to this phase as the current-CFL (curCFL) phase. 

In the present work we wish to show that the spacelike screening 
(Meissner) masses in the curCFL phase are real, and that the 
instability is indeed resolved. We also improve on the calculation
of \cite{Kryjevski:2005qq,Schafer:2005ym} in several ways. We 
allow the CFL gaps $\Delta_{1,2,3}$ to be different and we properly
implement electric charge neutrality. On the other hand, we make 
one important simplification as compared to our earlier work and
ignore homogeneous Goldstone boson condensates. This assumption
simplifies the calculation of the dispersion relations and allows 
a more detailed comparison with previous work on the gCFL phase. 
We will comment on the effect of kaon condensation in the conclusions. 
We also compare our results to studies of the LOFF phase in three
flavor QCD \cite{Alford:2000ze,Casalbuoni:2003wh,Casalbuoni:2005zp,Ciminale:2006sm,Mannarelli:2006fy}.
Recent work on Goldstone boson currents in two flavor QCD can be 
found in \cite{Giannakis:2004pf,Huang:2005pv,Hong:2005jv,Gorbar:2006up}.

%%%%%%%%%%%%%%%%%%%%%%%%%%%%%%%%%%%%%%%%%%%%%%%%%%%%%%%%%%%%%%%%%%%%%%%%%
\section{Effective Lagrangian and dispersion laws}
\label{sec_disp}
%%%%%%%%%%%%%%%%%%%%%%%%%%%%%%%%%%%%%%%%%%%%%%%%%%%%%%%%%%%%%%%%%%%%%%%%%

 We consider an effective lagrangian that describes the interaction 
of gapped fermions with background gauge fields
\cite{Kryjevski:2004jw,Kryjevski:2004kt,Kryjevski:2005qq,Schafer:2005ym}
\bea
\label{l_eff}
{\cal L } &=& 
     \Tr\left(\chi_L^\dagger(iv\cdot\partial -\hat{\mu}^L-A_e Q)\chi_L\right)
   + \Tr\left(\chi_R^\dagger(iv\cdot\partial -\hat{\mu}^R-A_e Q)\chi_R\right)
 \nonumber \\
 & & \mbox{}
  -i \Tr \left(\chi_L^\dagger \chi_L X v\cdot(\partial-iA^T)X^\dagger\right)
  -i \Tr \left(\chi_R^\dagger \chi_R Y v\cdot(\partial-iA^T)Y^\dagger\right)
 \nonumber \\
 & & \mbox{}
 -\frac{1}{2}\sum_{a,b,i,j,k} 
  \Delta_k\left(\chi_{L}^{ai}\chi_{L}^{bj}
                \epsilon_{kab}\epsilon_{kij}
               -\chi_{R}^{ai}\chi_{R}^{bj}
                \epsilon_{kab}\epsilon_{kij} + h.c. \right).
\eea
Here, $\chi_{L,R}^{ai}$ are left/right handed fermions with color 
index $a$ and flavor index $i$, $A_\mu$ are $SU(3)_C$ color gauge 
fields, and $\hat{\mu}^L=MM^\dagger/(2\mu)$, $\hat{\mu}^R=M^\dagger
M/(2\mu)$ are effective chemical potentials induced by the quark 
mass matrix $M$. The matrix $Q=\mathrm{diag}({2\over3},-{1\over3},
-{1\over3})$ is the quark charge matrix and $A_e$ is an electro-static 
potential. The fields $X,Y$ determine the flavor orientation of the 
left and right handed gap terms and transform as $X\to LXC^T$, $Y\to 
RYC^T$ under $(L,R)\in SU(3)_L \times SU(3)_R$ and $C\in SU(3)_C$, and 
$\Delta_k$ $(k=1,2,3)$ are the CFL gap parameters. The gauge field 
$A_\mu$ appears transposed because of the transformation properties
of $X,Y$. From the lagrangian equ.~(\ref{l_eff}) we can read off the 
Nambu-Gor'kov propagator (left handed part)
\be
\label{prop}
  \left(
  \begin{array}{cc} 
     G^+   & \Xi^- \\ 
     \Xi^+ & G^- \end{array}
  \right)=\left(
  \begin{array}{cc}
    (p_0-p){\bf1}+\mathcal{X}_{v}& \underline{\Delta} \\ 
    \underline{\Delta} & (p_0+p){\bf1}- \mathcal{X}^T_{-v}
  \end{array}
  \right)^{-1},
\ee
where $p=\vec v\cdot\vec p-\mu$, with the Fermi velocity $\vec v$.
The components of the propagator are matrices in color-flavor 
space. We use a basis spanned by the Gell-Mann matrices $\lambda^A$
$(A=1,\ldots,8)$ and $\lambda_0=\sqrt{2\over3}{\bf 1}$. In this basis
\be
  \underline{\Delta}^{AB}=-\textstyle{1\over2}\Delta_{ab}\varepsilon_{ija}
  \varepsilon_{rsb}\lambda^A_{ir} \lambda^B_{js}, 
  \qquad \Delta_{ab}=\mathrm{diag}(\Delta_1,\Delta_2,\Delta_3)_{ab}.
\ee
We will assume that $X=Y=1$ which excludes the possibility of kaon 
condensation. This assumption significantly simplifies the calculation 
of the fermion dispersion relations and the current correlation 
functions. It is possible to suppress kaon condensation by including 
a large instanton induced interaction \cite{Schafer:2002ty}. We 
will discuss instanton effects in Sect.~\ref{secinst}. The gauge 
field vertex in the phase with $X=Y=1$ is 
\be
  \mathcal{X}_{v,AB}={\textstyle{1\over2}}
  \Tr\left[\lambda_A(\hat\mu^L+A_e Q)\lambda_B+
  \lambda_A\lambda_B v\cdot A^T\right],
\ee
We define 
\be
\label{a0}
  A^{0T} = -\hat\mu^L-A_e Q+\tilde A_3\lambda_3+\tilde A_8\lambda_8,
\ee
where $\tilde A_3$, $\tilde A_8$ are the diagonal components of the
color gauge field relative to the leading order solution of the color 
neutrality condition. We wish to study the possibility of forming 
a Goldstone boson current. Gauge invariance implies that the free 
energy only depends on the combinations $\vec{J}_L=X^\dagger(\vec{\nabla}
-i\vec{A}^T)X$ and $\vec{J}_R=Y^\dagger(\vec{\nabla}-i\vec{A}^T)Y$. We 
will restrict ourselves to diagonal currents $\vec{J}_{L,R}$. Within 
our approximations the free energies of the vector and axial-vector 
currents $\vec{J}_L=\pm\vec{J}_R$ are degenerate. We will consider 
the pure vector current $\vec{J}_L=\vec{J}_R=\vec{A}^T$ with
\be
\label{cur}
  \vec{A}^{T}={\textstyle{1\over2}}\vec{\jmath}\,
   \left(\lambda_3+{\textstyle{1\over\sqrt{3}}}\lambda_8 
  -{\textstyle{1\over\sqrt{6}}}\lambda_0\right).
\ee
This ansatz has the feature that it does not shift the energy of electrically 
charged fermion modes. We will see that an ansatz of this type is favored 
if electric neutrality is enforced. The electric charge due to a gapless
charged fermion is $\delta Q\sim \mu^2\jmath$. The density of electrons 
is $n_e\sim \mu_e^3$ and electric neutrality gives $\mu_e\sim (\mu^2
\jmath)^{1/3}\gg \jmath$. This means that any solution with $\mu_e
\sim \jmath\sim \Delta$ has to have a very small density of gapless
charged fermions. We will study a more general ansatz for the current 
in Sec.~\ref{secgen} and show that the energetically preferred solution 
is indeed close to the ansatz equ.~(\ref{cur}). From the propagator 
(\ref{prop}) we obtain the following  dispersion laws,
\bea
\epsilon_1&=&{\tilde A_3\over2}-{\sqrt{3}\tilde A_8\over 2}-\mu_s
   -{\vec v\cdot\vec \jmath\over2}
   +\sqrt{\left(p+{\tilde A_3\over2}+{\tilde A_8\over 2\sqrt{3}}\right)^2
   +\Delta_1^2},
   \nonumber\\
\epsilon_2&=&-{\tilde A_3\over2}+{\sqrt{3}\tilde A_8\over 2}+\mu_s
   +{\vec v\cdot\vec \jmath\over2}
   +\sqrt{\left(p+{\tilde A_3\over2}+{\tilde A_8\over 2\sqrt{3}}\right)^2
   +\Delta_1^2},
   \nonumber\\
\epsilon_3&=&-{\tilde A_3\over2}-{\sqrt{3}\tilde A_8\over 2}+A_e-\mu_s
   +\sqrt{\left(p-{\tilde A_3\over2}+{\tilde A_8\over 2\sqrt{3}}
   -{\vec v\cdot\vec \jmath\over2}\right)^2+\Delta_2^2},
   \nonumber\\
\epsilon_4&=&{\tilde A_3\over2}+{\sqrt{3}\tilde A_8\over 2}-A_e+\mu_s
   +\sqrt{\left(p-{\tilde A_3\over2}+{\tilde A_8\over 2\sqrt{3}}
   -{\vec v\cdot\vec \jmath\over2}\right)^2+\Delta_2^2},
  \nonumber\\
\epsilon_5&=&\tilde A_3-A_e + \sqrt{\left(p-{\tilde A_8\over \sqrt{3}}
   -{\vec v\cdot\vec \jmath\over2}\right)^2+\Delta_3^2},
   \nonumber\\
\epsilon_6&=&-\tilde A_3+A_e +\sqrt{\left(p-{\tilde A_8\over \sqrt{3}}
   -{\vec v\cdot\vec \jmath\over2}\right)^2+\Delta_3^2},
   \nonumber\\
\epsilon_7&=&\sqrt{p^2+\Delta_1^2}+\ldots,\nonumber\\
\epsilon_8&=&\sqrt{p^2+\Delta_1^2}+\ldots,\nonumber\\
\epsilon_9&=&\sqrt{p^2+4\Delta_1^2}+\ldots.
\eea
The dots in $\epsilon_{7,8,9}$ denote 
terms of the order $a^n$ ($n\ge1$) with  $a\in\{\tilde 
A_3, \tilde A_8, \vec v\cdot\vec \jmath, \Delta_2-\Delta_1, \Delta_3-
\Delta_1\}$. We observe that without a current $\epsilon_1$ and 
$\epsilon_3$ become gapless at $\mu_s=\Delta$. We also note that 
the current shifts the energy of the electrically neutral modes
$\epsilon_{1,2}$ and the momenta of the charged modes $\epsilon_{3-6}$.

%%%%%%%%%%%%%%%%%%%%%%%%%%%%%%%%%%%%%%%%%%%%%%%%%%%%%%%%%%%%%%%%%%%%%%%%%
\section{Free energy}
\label{sec_f}
%%%%%%%%%%%%%%%%%%%%%%%%%%%%%%%%%%%%%%%%%%%%%%%%%%%%%%%%%%%%%%%%%%%%%%%%%

 In order to determine the groundstate we need to supplement the 
effective lagrangian with a potential for the order parameter $\Delta_k$.
In QCD this potential is determined by gauge field dynamics. For the 
present purpose we are not interested in the details of this interaction
and we will model the effective potential as a quadratic function 
of $\Delta_k$. The free energy is given by
\be
\label{om1}
  \Omega={1\over G}(\Delta_1^2+\Delta_2^2+\Delta_3^2)
   -{1\over G_i} \sqrt{\mu_s}\Delta_3^2
   +{3\mu^2 \jmath^2\over8\pi^2}-{\mu^2\over4\pi^2}
  \int dp\int dt 
  \sum_{i=1}^9(|\epsilon_i|-|p|), 
\ee
with $t=\cos\theta$. Here, $G$ fixes the magnitude of the gap 
in the chiral limit and $G_i$ takes into account instanton effects. 
The structure of the instanton term is determined by vacuum 
expectation value of the 't Hooft interaction. The value of $G_i$ 
can be computed in perturbative QCD \cite{Schafer:2002ty} but in 
the present work we will treat it as a free parameter. We first 
study the case that there are no instanton effects, $1/G_i=0$. The 
integral in equ.~(\ref{om1}) is quite complicated. We assume that 
the external fields and currents are small and expand in a set of 
parameters $b\in\{\tilde A_3/\Delta_1, \tilde A_8/\Delta_1, \jmath/\Delta_1, 
(\Delta_2-\Delta_1)/\Delta_1, (\Delta_3-\Delta_1)/\Delta_1, (\mu_s-\Delta_1)
/\Delta_1\}$. We will study the convergence properties of this expansion 
below. To second order in $b$ we obtain
\bea
\label{om2}
\Omega  & \simeq &
   {1\over G}\left(\Delta_1^2+\Delta_2^2+\Delta_3^2\right)
   +{\mu^2\over54\pi^2}
   \Big[-54\left(\Delta_1^2+\Delta_2^2+\Delta_3^2\right)
   \log\left({\mu\over\Delta_1}\right)
      \nonumber\\
  &&\mbox{}
     +   \Delta_1^2 (15-34\log2)
     + 2(\Delta_2^2+\Delta_3^2)(21-17\log2)
      \nonumber\\
  &&\mbox{}
     - 16\Delta_1\left(\Delta_2+\Delta_3\right)
       (6-\log2)+4\Delta_2\Delta_3(3+4\log2)\Big]
      \nonumber\\
  &&\mbox{}
    +{\mu^2\jmath^2\over648\pi^2}(219-32\log2)
    +{2\mu^2\over9\pi^2}\left(\tilde A_3^2+\tilde A_8^2\right)(-3+2\log2)
      \nonumber\\
  &&\mbox{}
    +H\left(\Delta_1,-{\tilde A_3\over2}
    -{\tilde A_8\over2\sqrt{3}},-{\tilde A_3\over2}
    +{\sqrt{3}\tilde A_8\over2}+\mu_s,0,\jmath\right) 
      \nonumber\\
  &&\mbox{}
    +H\left(\Delta_2,{\tilde A_3\over2}
    -{\tilde A_8\over2\sqrt{3}},{\tilde A_3\over2}
    +{\sqrt{3}\tilde A_8\over2}-A_e+\mu_s,\jmath,0\right)
      \nonumber\\
  &&\mbox{}
    +H\left(\Delta_3,{\tilde A_8\over\sqrt{3}},0,\jmath,0\right),
\eea
where the function $H$ is given by
\bea
\label{om3}
 H(\Delta,a_1,a_2,\jmath_1,\jmath_2)
  &=& -{\mu^2\over12\pi^2}
         \left[6\Delta^2\left(1+2\log\left({2\mu\over\Delta}\right)\right)
      + 12a_1^2+\jmath_1^2\right]
      \nonumber\\
  & & \hspace{0.23cm}\mbox{}
      +{2\mu^2\sqrt{\Delta}\over15\pi^2\jmath_2}
    \Big[(2a_2-2\Delta-\jmath_2)^{5/2}\Theta(2a_2-2\Delta-\jmath_2)
     \nonumber\\
  & & \hspace{1.65cm}\mbox{}
    -(2a_2-2\Delta+\jmath_2)^{5/2}\Theta(2a_2-2\Delta+\jmath_2)\Big].
\eea
The gap equations and neutrality equations can be summarized as $E_i=0$, 
where 
\bea
E_i&=&\frac{\partial\Omega}{\partial\Delta_i}, \hspace{0.25cm}
(i=1,2,3) \\
E_4&=&\frac{\partial\Omega}{\partial A_e}, \hspace{0.5cm}
E_5 = \frac{\partial\Omega}{\partial\tilde A_3},\hspace{0.25cm}
E_6 = \frac{\partial\Omega}{\partial\tilde A_8}. 
\eea
From the linear combination $E_4+E_5+{1\over\sqrt{3}}E_6$ we obtain 
the relation
\be
  \tilde A_8=-\sqrt{3}\tilde A_3
\ee
Including the $(-A_e^4)$ term in $\Omega$, we find from $E_4$ 
(for $\mu\gg A_e$)
\be
  A_e=\mathrm{max}(0,\mu_s-\Delta_2-\tilde A_3).
\ee
We assume $\mu_s-\Delta_2-\tilde A_3<0$ (and therefore $A_e=0$), which 
will turn out to be consistent for the values of $\mu_s$ under consideration.
Then we find $\Delta_3=\Delta_2$. Furthermore, $E_2$ is then a linear 
equation in $\Delta_2$, with the solution
\be
  \Delta_2=\Delta_1\left(1-{54y\over51+54y-4\log2}\right),
\ee
where $y=\log(\Delta_1/\Delta(0))$ and $\Delta(0)$ is the value of the 
gap in the chiral limit.
The linear combination $E_1-E_4-2E_5$ 
gives a linear equation for $\tilde A_3$, with the solution
\begin{equation}
  \tilde A_3=-\Delta_1{54y\over21-8\log2}
   \left(1-{36(1+y)\over51+54y-4\log2}\right).
\end{equation}
There remains one equation (proportional to $E_4+2E_5$), 
\begin{equation}
  {\mu^2(21-8\log 2)\over18\pi^2}\tilde A_3
   -{\mu^2\sqrt{\Delta_1}\over 6\pi^2\jmath}
  \left[\Theta(s_+)s_+^{3/2}-\Theta(s_-)s_-^{3/2}\right]=0, \label{e14}
\end{equation}
where $s_\pm=2\mu_s-2\Delta_1-4\tilde A_3\pm \jmath$. We solve equ. (\ref{e14})
numerically for $\Delta_1$ for given values of $\mu_s$ and $\jmath$. 

%%%%%%%%%%%%%%%%%%%%%%%%%%%%%%%%%%%%%%%%%%%%%%%%%%%%%%%%%%%%%%%%%%%%%%%%%
\begin{figure}
\includegraphics[width=7cm]{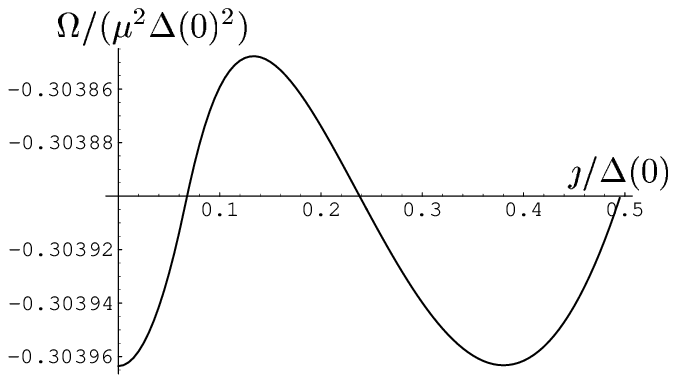}
\includegraphics[width=8cm]{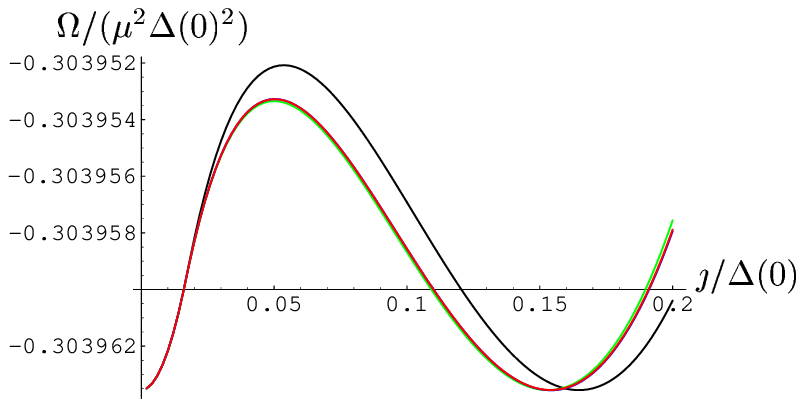}
\caption{Left panel: $\Omega/(\mu^2\Delta(0)^2)$ as a function of $\jmath/
\Delta(0)$ for $\Delta_1=\Delta_2=\Delta_3=\Delta(0)$, $A_e=\tilde A_3=
\tilde A_8=0$ at $\mu_s=0.9683\Delta(0)$ ($O(b^2)$ approximation). Right 
panel: $\Omega/(\mu^2\Delta(0)^2)$ as a function of $\jmath/\Delta(0)$ with 
gap equations and neutrality conditions taken into account. The different 
curves correspond to different orders in the expansion in $b$: second order 
(black), third order (green), fourth order (blue), fifth order (red). All 
curves are shown for the respective values of $\mu_{s,crit}$.
\label{figomj}}
\end{figure}
%%%%%%%%%%%%%%%%%%%%%%%%%%%%%%%%%%%%%%%%%%%%%%%%%%%%%%%%%%%%%%%%%%%%%%%%%

 We observe that $\Omega/(\mu^2\Delta(0)^2)$ is a function of $b/\Delta(0)$
and $\Delta_1/\Delta(0)$. This means that in order to study the effective 
potential as a function of dimensionless variables we do not have to 
specify the value of $\mu$ and $G$. We first study the dependence of 
the thermodynamic potential $\Omega$ on the current $\jmath$ for different 
$\mu_s$. We begin with the simplified case $\Delta_1=\Delta_2=
\Delta_3=\Delta(0)$, $A_e=\tilde A_3=\tilde A_8=0$. We find that $\Omega(
\jmath)$ develops a nontrivial minimum for $\mu_s>0.9683\Delta(0)$, as 
shown in the left panel of Fig.~\ref{figomj}. If we take into account the 
gap equations and neutrality conditions, a nontrivial minimum appears 
for a slightly larger value of $\mu_s$, namely $\mu_s>\mu_{s,crit}\equiv
0.9916\Delta(0)$, as shown in the right panel of Fig.~\ref{figomj} (black 
curve). We note that the expectation value of $\jmath$ is considerably 
smaller than in the simplified case.

%%%%%%%%%%%%%%%%%%%%%%%%%%%%%%%%%%%%%%%%%%%%%%%%%%%%%%%%%%%%%%%%%%%%%%%%%
\begin{figure}
\includegraphics[width=7cm]{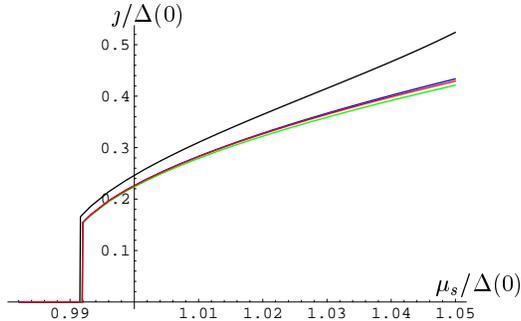}
\caption{Optimum current $\jmath/\Delta(0)$ as a function of $\mu_s/\Delta
(0)$. We compare results at different orders in the small parameters $b$
(curves labeled as in Fig.~\ref{figomj}).
\label{figjmus}}
\end{figure}
%%%%%%%%%%%%%%%%%%%%%%%%%%%%%%%%%%%%%%%%%%%%%%%%%%%%%%%%%%%%%%%%%%%%%%%%%

 We have also studied the importance of higher order terms in $b$
in the free energy. These terms are not included in 
equ.~(\ref{om2},\ref{om3}) but they can easily be generated using 
a symbolic algebra package such as Mathematica or Maple. In 
Fig.~\ref{figomj} we compare the $O(b^2)$ result to the $O(b^n)$
approximation with $n=3,4,5$. At the minimum of $\Omega$ we find 
$\Delta_3=\Delta_2$ and $A_e=0$ as before. We also see that while 
the second order approximation gives results that are only qualitatively 
correct, the differences between the results of the higher order 
approximations are very small.

We may also determine the minimum of the free energy for $\mu_s$ greater than 
the critical value. Fig.~\ref{figjmus} shows the expectation value of $\jmath$ 
as a function of $\mu_s$. Again, we compare results at different order in 
the small parameter $b$. We can see that the differences between the results 
of the higher order approximations are very small even for (somewhat) higher 
values of $\mu_s$. The highest approximation we discuss is the fifth order 
approximation. In that case we find that the critical values of $\mu_s$ is 
given by $\mu_{s,crit}=0.9919 \Delta(0)$. The expectation value of the 
$\jmath$ is equal to $0.1541 \Delta(0)$ at this point.

\begin{figure}
\includegraphics[width=8cm]{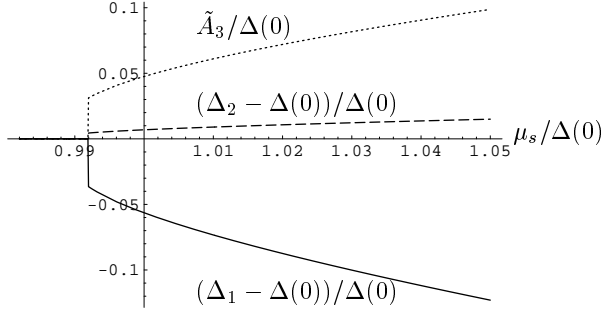}  
\caption{Gauge field $\tilde A_3/\Delta(0)$ (dotted) and flavor 
symmetry breaking in the gap parameters $(\Delta_1-\Delta(0))/\Delta(0)$ 
(continuous) and $(\Delta_2-\Delta(0))/\Delta(0)$ (dashed) as functions 
of $\mu_s/\Delta(0)$, computed to $O(b^5)$.\label{figadjmus}}
\end{figure}

 In Fig.~\ref{figadjmus} we show the gauge potential $\tilde A_3$ and the 
flavor asymmetries in the gaps, $\Delta_1-\Delta(0)$, $\Delta_2-\Delta(0)$, 
as a function of $\mu_s$ calculated to $O(b^5)$. We remark that $\epsilon_1$ 
becomes a gapless mode for $\mu_s>\mu_{s,crit}$, while $\epsilon_3$ (the 
almost quadratic mode of Ref.~\cite{Alford:2004hz}) has at least a very 
small gap, greater than $0.04 \Delta(0)$, for all values of $\mu_s$. 

%%%%%%%%%%%%%%%%%%%%%%%%%%%%%%%%%%%%%%%%%%%%%%%%%%%%%%%%%%%%%%%%%%%%%%%%%
%%%%%%%%%%%%%%%%%%%%%%%%%%%%%%%%%%%%%%%%%%%%%%%%%%%%%%%%%%%%%%%%%%%%%%%%%

%%%%%%%%%%%%%%%%%%%%%%%%%%%%%%%%%%%%%%%%%%%%%%%%%%%%%%%%%%%%%%%%%%%%%%%%%  
\section{Meissner masses}
\label{sec_meiss}
%%%%%%%%%%%%%%%%%%%%%%%%%%%%%%%%%%%%%%%%%%%%%%%%%%%%%%%%%%%%%%%%%%%%%%%%%

 In this section we study the stability of the current-current
correlation functions. The screening (Meissner) masses are given by
\bea
  (m^2_M)_{ab}^{ij} &=& {g^2\mu^2\over2\pi^2} \delta_{ab}\delta^{ij}
   + {g^2\mu^2 \over16\pi^2}\lim_{k\to0}\lim_{k_0\to0}
   \int dp\int dt\int {dp_0\over{2\pi}}
       \nonumber \\[0.2cm]
  & &   \hspace{2cm}\mbox{} \times\mathrm{Tr} 
      \Big[ G^+(p) V_a^i G^+(p+k)V_b^j  
         +  G^-(p)\tilde V_a^i G^-(p+k)\tilde V_b^j \nonumber \\
  & & \hspace{2.7cm}\mbox{}
         + \Xi^+(p) V_a^i\Xi^-(p+k)\tilde V_b^j
         + \Xi^-(p)\tilde V_a^i \Xi^+(p+k) V_b^j\Big],
\eea
where we have defined the vertices 
\begin{equation}
  (V_a^i)_{AB}=\textstyle{1\over2}\Tr[\lambda_A\lambda_B\lambda_a^T]v^i, 
   \qquad (\tilde V_a^i)_{AB}=-\textstyle{1\over2}
   \Tr[\lambda_B\lambda_A\lambda_a^T] \tilde v^i,
\end{equation}
with $\tilde v^i=-v^i$.
We begin by studying the screening masses for small $\mu_s<\mu_{s,crit}$.
In this case the Meissner mass matrix is diagonal, and we have $m_{M11}^2
=m_{M22}^2=m_{M33}^2=m_{M88}^2$ and $m_{M44}^2=m_{M55}^2
=m_{M66}^2=m_{M77}^2$, with
\bea
  m_{M11}^2 &=& g^2\mu^2 \, {21-8\log2\over108\pi^2},\\
  m_{M44}^2 &=& {g^2\mu^2\over\pi^2}
                 \bigg[-{\log2-3\alpha^2\over6\alpha^2}
             -{1\over\alpha\beta}\arctan\left({\alpha\over\beta}\right)
       +{1-\alpha^2\over2\alpha^2\gamma}\mathrm{arctanh}
      \left({\gamma\over1+\beta^2}\right)\bigg],
\eea
with $\alpha=\mu_s/\Delta(0)$, $\beta=\sqrt{4-\alpha^2}$ and $\gamma=
\sqrt{9-10\alpha^2+\alpha^4}$. We note that the Meissner masses of 
the modes $i=1,2,3,8$ are independent of $\mu_s$ below $\mu_{s,crit}$.
These modes only couple to pairs of quasi-particles that are subject 
to the same shift in energy. In our approximation, in which we neglect 
the effect of $\mu_s$ on the density of states, this implies that 
the $\mu_s $ dependence can be eliminated by a simple shift in the 
loop energy. This is not the case for the off-diagonal gluon modes
$i=4,5,6,7$. We find that the screening masses $m_{Mii}$ ($i=4,5,6,7$) 
drop with $\mu_s$, but they remain real and non-zero for $\mu_s<
\mu_{s,crit}$  \cite{Casalbuoni:2004tb}. 

%%%%%%%%%%%%%%%%%%%%%%%%%%%%%%%%%%%%%%%%%%%%%%%%%%%%%%%%%%%%%%%%%%%%%%%%%
\begin{figure}
\includegraphics[width=7cm]{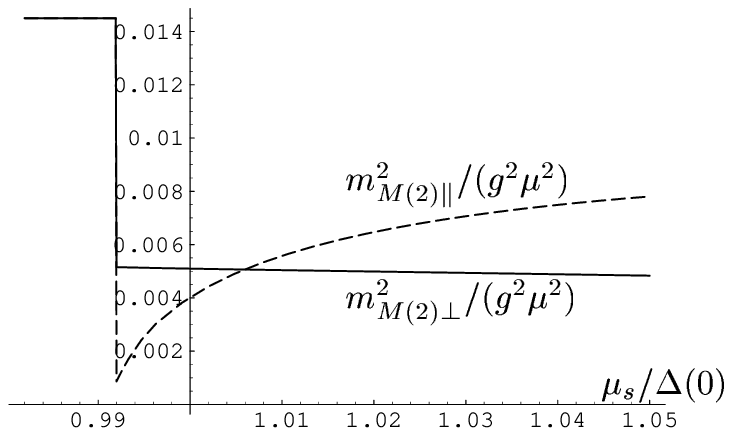}
\includegraphics[width=7cm]{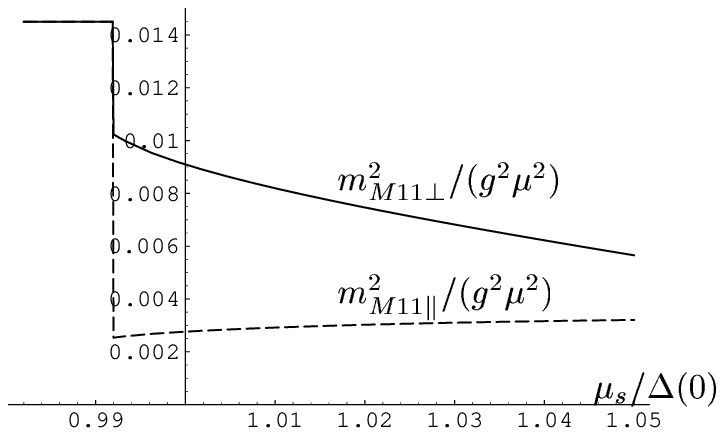}
\caption{Meissner masses squared $m_{M(2)\perp}^2/(g^2\mu^2)$, 
$m_{M(2)\parallel}^2/(g^2\mu^2)$, $m_{M11\perp}^2/(g^2\mu^2)$ and 
$m_{M11\parallel}^2/(g^2\mu^2)$ as functions of $\mu_s/\Delta(0)$, 
computed to $O(b^5)$.  \label{figmh}}
\end{figure}
%%%%%%%%%%%%%%%%%%%%%%%%%%%%%%%%%%%%%%%%%%%%%%%%%%%%%%%%%%%%%%%%%%%%%%%%%

In the presence of a finite current $\vec \jmath$ we may decompose the 
Meissner masses into a longitudinal and a transverse component,
\be
  (m_M^2)^{ij}=m_{M\perp}^2(\delta^{ij}-\hat \jmath^i\hat \jmath^j)
   +m_{M\parallel}^2\hat \jmath^i\hat \jmath^j. 
\ee
First let us evaluate the Meissner masses through order $b^0$, which 
corresponds to the $O(b^2)$-approximation in the free energy.
We will see that the interesting terms are those of order $b^{-1/2}$.

A chromomagnetic instability could occur for the Meissner masses with 
color indices 1,2,3 or 8 \cite{Casalbuoni:2004tb}. In the 3-8-sector 
there are contributions where both quark propagators in the gluon self 
energy diagram are gapless. We find that the mass matrix in the 3-8-sector 
is not diagonal, the mixing angle being equal to $\pi\over6$. We obtain 
the following eigenvalues of the 3-8-mass matrix within our approximations,
\bea
  m_{M(1)\perp,\parallel}^2&=&g^2\mu^2{21-8\log2\over108\pi^2},\\
  m_{M(2)\perp}^2&=&g^2\mu^2\bigg\{{21-8\log2\over108\pi^2}
    +{\sqrt{\Delta_1}\over45\pi^2\jmath^3}\Big[\Theta(s_+)s_+^{3/2}
     (2s_+\!-5\jmath) \nonumber\\
 & & \hspace{0.5cm}\mbox{}
   -\Theta(s_-)s_-^{3/2}(2s_-\!+5\jmath)\Big] \bigg\},\\
   m_{M(2)\parallel}^2&=&g^2\mu^2\bigg\{{21-8\log2\over108\pi^2}
  -{\sqrt{\Delta_1}\over90\pi^2\jmath^3}\big[\Theta(s_+)
   \sqrt{s_+}(15\jmath^2-20\jmath s_++8s_+^2)\nonumber\\ 
  & & \hspace{0.5cm}\mbox{}
    -\Theta(s_-)\sqrt{s_-}(15\jmath^2+20\jmath s_-+8s_-^2)\big] \bigg\}.
\eea
These quantities are positive for $\mu_s=\mu_{s,crit}$ when $\jmath$ is 
chosen such that $\Omega$ is minimal. 
We remark that the longitudinal Meissner mass squared $m_{M(2)\parallel}^2$ 
is very similar to the second (partial) derivative of $\Omega$ with respect 
to $\jmath$ (only the $\Theta$-independent terms are different because 
the current contains a component proportional to $\lambda_0$.) 

For $m_{M11}^2(=m_{M22}^2)$ there are contributions where one quark propagator 
in the gluon self energy diagram is gapless and the other one contains the the 
would-be quadratic mode. We find
\bea
  m_{M11\perp}^2&=&\!g^2\mu^2\Bigg\{{21-8\log2\over108\pi^2}
   -{\sqrt{\Delta_1}\over60\pi^2\jmath^3}\bigg[\Theta(s_+)
    \bigg(-\sqrt{s_+}\big(8s_+^2+15r^2+25rs_+\nonumber\\
  & & \hspace{0.5cm}\mbox{}
     -10\jmath(2s_++3r)\big)+15\sqrt{r}\left((s_++r-\jmath)^2-\jmath^2\right)
      \arctan\sqrt{s_+\over r}\bigg)\nonumber\\
  & & \hspace{0.5cm}\mbox{}
     -\left(\jmath\leftrightarrow -\jmath,\ 
      s_+\leftrightarrow s_-\right)\bigg]\Bigg\}, \\
m_{M11\parallel}^2&=&g^2\mu^2\Bigg\{{21-8\log2\over108\pi^2}
  -{\sqrt{\Delta_1}\over30\pi^2\jmath^3}\bigg[\Theta(s_+)
  \bigg(\sqrt{s_+}\big(8s_+^2+15r^2+25rs_+\nonumber\\
  & & \hspace{0.5cm}\mbox{}
     -5\jmath(4s_++6r-3\jmath)\big)-15
  \sqrt{r}(s_++r-\jmath)^2\arctan\sqrt{s_+\over r}\bigg)\nonumber\\
  & & \hspace{0.5cm}\mbox{}
      -\left(\jmath\leftrightarrow -\jmath,\ s_
    +\leftrightarrow s_-\right)\bigg]\Bigg\},
\eea
where $r=2(\Delta_2-\mu_s+\tilde A_3)>0$. Again we find that $m_{M11\perp}^2$ 
and $m_{M11\parallel}^2$ are positive.

The squared Meissner masses $m_{M44}^2$ to $m_{M77}^2$ are equal at 
$O(b^0)$. At the above value of $\mu_s$ we find
\be
  m_{M44}^2={9-\sqrt{3}\pi-3\log2\over18\pi^2},
\ee
which is again positive.

As above we may include higher orders in $b$ in the calculation of the 
Meissner masses. Fig.~\ref{figmh} shows the ``dangerous'' components of 
the Meissner masses as functions of $\mu_s$ at $O(b^5)$. 
We observe that all masses are real.
As above the differences between the results of the 
higher order approximations are very small. The numerical values of 
the ``dangerous'' components of the Meissner mass just above the phase 
transition are shown in the first line of Table \ref{tabm}.

%%%%%%%%%%%%%%%%%%%%%%%%%%%%%%%%%%%%%%%%%%%%%%%%%%%%%%%%%%%%%%%%%%%%%%%%%
\section{An improved ansatz for the current}
\label{secgen}
%%%%%%%%%%%%%%%%%%%%%%%%%%%%%%%%%%%%%%%%%%%%%%%%%%%%%%%%%%%%%%%%%%%%%%%%%

 We have studied the effect of using a more general ansatz for the 
current. Consider 
\be
  \vec{A}^{T}={\textstyle{1\over2}}\vec{\jmath}\,\left(c_1\lambda_3
  +c_2{\textstyle{1\over\sqrt{3}}}\lambda_8
  -c_3{\textstyle{1\over\sqrt{6}}}\lambda_0\right) \label{jgen}
\ee
The ansatz given in equ.~(\ref{cur}) corresponds to $c_1=c_2=c_3=1$. 
We now minimize $\Omega$ with respect to $\jmath$ and the coefficients 
$c_i$. We can fix $c_1$ such that the explicit $\jmath^2$ term in 
equ.~(\ref{om1}) remains unchanged as $c_2$ and $c_3$ are varied, 
which amounts to setting $c_1=\sqrt{(9-2c_2^2-c_3^2)/6}$. At 
$O(b^5)$ we find $\mu_{s,crit}=0.9918\Delta(0)$, and
\be
  c_1=0.980, \quad c_2=1.000, \quad c_3=1.114. \label{csol}
\ee
We observe that the exact minimum is indeed quite close to our
simple ansatz. The ``dangerous'' components of the Meissner mass 
are not significantly affected in the vicinity of the phase transition. 
Their values just above the phase transition are shown in the second 
line of Table \ref{tabm}.

%%%%%%%%%%%%%%%%%%%%%%%%%%%%%%%%%%%%%%%%%%%%%%%%%%%%%%%%%%%%%%%%%%%%%%%%%
\begin{table}
\begin {tabular}{|c||c|c|c|c|}\hline
% & $m_{M(2)\perp}^2$  & $m_{M(2)\parallel}^2$ & $m_{M11\perp}^2$ 
%                      & $m_{M11\parallel}^2$\\ \hline\hline
 & $m_{M(2)\perp}^2/m_M^2(0)$  & $m_{M(2)\parallel}^2/m_M^2(0)$ 
 & $m_{M11\perp}^2/m_M^2(0)$   & $m_{M11\parallel}^2/m_M^2(0)$\\ \hline\hline
 $c_i=1$ & 
 %   $0.00516$ & $0.00082$ & $0.01025$ & $0.00253$ \\ \hline
     $0.356$   & $0.056$   & $0.707$   & $0.174$ \\ \hline
 $c_{1,3}\neq1$ & 
 %   $0.00519$ & $0.00089$ & $0.01009$ & $0.00207$ \\ \hline
     $0.358$   & $0.061$   & $0.696$   & $0.142$ \\ \hline
\end{tabular}
\caption{Dangerous components of the Meissner masses at $\mu_s=\mu_{s,crit}$ 
in units of the screening mass at $\mu_s=0$ (third order approximation). 
First line: $c_{1,2,3}=1$ in equ.~(\ref{jgen}), second line: $c_i$ 
according to equ.~(\ref{csol}). 
\label{tabm}}
\end{table}
%%%%%%%%%%%%%%%%%%%%%%%%%%%%%%%%%%%%%%%%%%%%%%%%%%%%%%%%%%%%%%%%%%%%%%%%%

%%%%%%%%%%%%%%%%%%%%%%%%%%%%%%%%%%%%%%%%%%%%%%%%%%%%%%%%%%%%%%%%%%%%%%%%%
\section{Instanton contribution}
\label{secinst}
%%%%%%%%%%%%%%%%%%%%%%%%%%%%%%%%%%%%%%%%%%%%%%%%%%%%%%%%%%%%%%%%%%%%%%%%%

 In this section we consider the role of an instanton induced
interaction. The value of $G_i$ can be computed in perturbative 
QCD \cite{Schafer:2002ty} but the result is very sensitive to 
the QCD scale parameter and to higher order perturbative corrections.
In the present work we have neglected kaon condensation and we
will therefore explore the effect of an instanton term of the 
magnitude required to suppress kaon condensation. We have previously 
argued \cite{Schafer:2002ty} that an instanton terms of this size is 
problematic from the point of view of instanton phenomenology at zero 
baryon density, but we will ignore this issue here. 

\begin{figure}
\includegraphics[width=7cm]{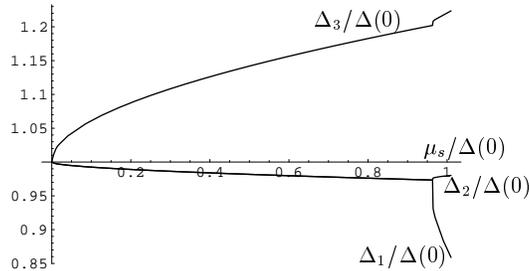}
\caption{$\Delta_i/\Delta(0)$ as functions of $\mu_s/\Delta(0)$
with the instanton contribution taken into account ($1/G_i=0.67
f_\pi^2\Delta(0)^{-1/2}$), computed to $O(b^5)$. Below 
the phase transition $\Delta_1$ and $\Delta_2$ are identical.
\label{figdeltai}}
\end{figure}

In order to suppress kaon condensation, $1/G_i$ has to
satisfy
\be
  {1\over G_i}>{f_\pi^2\mu_s^{3/2}\over\Delta_3^2}, \label{c1} 
\ee
where $f_\pi$ is the pion decay constant in the CFL phase. We have 
used $1/G_i=0.67f_\pi^2\Delta(0)^{-1/2}$ together with the perturbative 
result for the pion decay constant \cite{Son:1999cm}
\be 
 f_\pi^2 = \frac{21-8\log(2)}{18}\left(\frac{\mu^2}{2\pi^2}\right).
\ee
In this case the effective potential $\Omega(\jmath)$ develops a
non-trivial minimum for $\mu_s>\mu_{s,crit}=0.964\Delta(0)$ (using 
$\Omega$ to $O(b^5)$ and $c_i=1$). At this point we find 
$\Delta_3=1.21\Delta(0)$, such that the condition in equ.~(\ref{c1}) 
is just met. The magnitude of the current at the onset is $\jmath=
0.17\Delta(0)$ and the shape of the effective potential is not 
strongly modified by instanton effects. We also find that all 
Meissner masses are real. The main effect of the instanton term 
is a non-zero splitting between $\Delta_3$ and $\Delta_{1,2}$ below
$\mu_{s,crit}$, as shown in Fig.~\ref{figdeltai}. This effect leads
to the small reduction in the critical value of the effective 
chemical potential, $\mu_{s,crit}=0.964\Delta(0)$ compared to 
$\mu_{s,crit}=0.992\Delta(0)$ without the instanton term. We 
also note that because of the larger splitting between the gaps 
our expansion converges more slowly. 

%%%%%%%%%%%%%%%%%%%%%%%%%%%%%%%%%%%%%%%%%%%%%%%%%%%%%%%%%%%%%%%%%%%%%%%%%
%%%%%%%%%%%%%%%%%%%%%%%%%%%%%%%%%%%%%%%%%%%%%%%%%%%%%%%%%%%%%%%%%%%%%%%%%

%%%%%%%%%%%%%%%%%%%%%%%%%%%%%%%%%%%%%%%%%%%%%%%%%%%%%%%%%%%%%%%%%%%%%%%%%
\section{Conclusions}
\label{sec_sum}
%%%%%%%%%%%%%%%%%%%%%%%%%%%%%%%%%%%%%%%%%%%%%%%%%%%%%%%%%%%%%%%%%%%%%%%%%

 We have shown that near the point at which gapless fermion modes 
appear in the spectrum the CFL phase becomes unstable with respect 
to the formation of a Goldstone boson current. We have computed 
the magnetic screening masses in the Goldstone current phase and shown 
that all masses are real. Our calculation is based on an effective 
lagrangian of fermions coupled to a mean-field gap term and background
gauge potentials. The CFL pairing energy is of order $\mu^2\Delta^2$. 
We have performed an expansion in powers of $((\mu_s-\Delta)/\Delta,A/\Delta,
\jmath/\Delta,\delta\Delta/\Delta)$ and included all orders in this 
expansion until numerical convergence was obtained. We have neglected
all contributions that are suppressed by extra factors of $\Delta/\mu$.

 In this paper we have not included the effect of a homogeneous kaon 
condensate. Kaon condensation is likely to play an important role 
since the kaon condensate modifies the spectrum of fermions in such 
a way that the lightest mode is charged. We have considered the 
possibility that the condensate is suppressed by a large instanton
term. We find that the instanton term leads to a splitting between
the different CFL gaps for $\mu_s<\mu_{s,crit}$ but it does not 
qualitatively change the effective potential for the Goldstone 
boson current. 

 The phase studied in this paper is equivalent to a single plane wave
LOFF state. The constant gauge field $\vec A$ can be removed by a 
gauge transformation which leads to the following gap matrix 
(for $c_i=1$),
\be
  \Delta_{ab}=\mathrm{diag}
  (\Delta_1 e^{-i \vec x\cdot\vec\jmath},\Delta_2,\Delta_2)_{ab}.
\ee
This gap matrix is different from the ansatz $\Delta_{ab}= 
\mathrm{diag}(0,\Delta_2,\Delta_2)_{ab}e^{-i \vec x\cdot\vec\jmath}$
considered by Casalbuoni et al.~in \cite{Casalbuoni:2005zp}. One 
important difference is that we consider the CFL phase near the 
onset of the Goldstone boson current instability. In that regime 
all fermions remain paired and the free energy can be computed 
as an expansion in $(\jmath/\Delta)$. Casalbuoni et al.~consider 
the CFL phase near the maximum value of $\mu_s$ at which pairing 
between different flavors is possible and expand the free energy 
in $(\Delta/\jmath)$. Clearly, it is important to understand how
to interpolate between these two extremes. Once a kaon condensate 
develops the current is no longer diagonal, and it is also not 
equivalent to a pure gauge field. 

Acknowledgments: We would like to thank A.~Kryjevski for useful
discussions. This work is supported in part by the US Department 
of Energy grant DE-FG-88ER40388.

\end{document}